\documentclass[final, conference]{IEEEtran}
\usepackage[utf8]{inputenc}
\usepackage{graphicx}
\usepackage[noadjust]{cite}
\usepackage{multirow}
\usepackage{mathtools}
\usepackage[caption=false]{subfig}
\usepackage{balance}
\usepackage[ 
  bookmarks = true,
  bookmarksopen = true,
  bookmarksnumbered = true,
  breaklinks = true,
  linktocpage,
  pagebackref = false,
  colorlinks = true,
  linkcolor = blue,
  urlcolor  = blue,
  citecolor = red,
  anchorcolor = green,
  hyperindex = true,
  hyperfigures
  ]{hyperref}
\usepackage{url}

\title{Deseeding Energy Consumption of Network Stacks}

\author{
  \authorblockN{Iñaki Ucar\authorrefmark{1} and Arturo Azcorra\authorrefmark{1}\authorrefmark{2}}

  \authorblockA{
    \authorrefmark{1}Dept. of Telematics Engineering, Universidad Carlos III de Madrid, Spain\\
    \authorrefmark{2}IMDEA Networks Institute, Madrid, Spain\\
    \texttt{inaki.ucar@uc3m.es}, \texttt{azcorra@it.uc3m.es}
  }
}

\begin{document}

\maketitle

\begin{abstract}

Regular works on energy efficiency strategies for wireless communications are based on classical energy models that account for the wireless card only. Nevertheless, there is a non-negligible energy toll called \emph{cross-factor} that encompasses the energy drained while a frame crosses the network stack of an OS.

This paper addresses the challenge of deepen into the roots of the cross-factor, deseed its components and analyse its causes. Energy issues are critical for IoT devices. Thus, this paper conceives and validates a new comprehensive framework that enables us to measure a wide range of wireless devices, as well as multiple devices synchronously. We also present a rigorous methodology to perform whole-device energy measurements in laptops, a more generic and suitable device to perform energy debugging. Finally, and using this framework, we provide a collection of measurements and insights that deepens our understanding of the cross-factor.

\end{abstract}


\let\thefootnote\relax\footnote{
 \\\textcopyright 2015 IEEE. Personal use of this material is permitted. Permission from IEEE must be obtained for all other uses, in any current or future media, including reprinting/republishing this material for advertising or promotional purposes, creating new collective works, for resale or redistribution to servers or lists, or reuse of any copyrighted component of this work in other works.\\

 \noindent DOI: \hyperref{http://dx.doi.org/10.1109/RTSI.2015.7325085}{}{}{10.1109/RTSI.2015.7325085}
}

\section{Introduction}\label{sec:intro}

We are living in an era in which consumer electronics are becoming \emph{wireless consumer electronics}. That is, just about any known gadget is capable of connecting to cellular, wireless LAN or wireless PAN networks nowadays. The Internet of Things (IoT) is growing fast and wireless communications become its main driver. Due to the densification of wireless networks and the ubiquity of battery-powered devices, energy efficiency stands as a major research issue in the IoT.

More and more devices around us are becoming \emph{smart}, incorporating more processing power in order to do more things, and some of them are already comparable to a small laptop computer. As a consequence, not only do they share hardware components, but also software: in particular, the Linux kernel is spreading into billions of devices all over the world, whether inside of the popular Android operating system or other embedded systems.

Whereas a lot of research were and is devoted to obtain more energy-efficient hardware components (e.g., wireless cards, processors, screens), too little attention has been paid, in terms of energy, to a core software component that enables all these devices: the operating system and, inside it, the network stack. A recent study \cite{Serrano2014} unveils that the energy toll ---called \emph{cross-factor}--- ascribed to a frame crossing the operating system (i.e., the network stack and the wireless driver) is anything but negligible. In fact, it may account up to 97\% of the total energy consumption per frame in some devices. This poses doubts on prior works that propose energy efficiency strategies taking into account the wireless card only.

Therefore, this inspired us to deepen into the roots of the cross-factor, deseed its components and analyse its causes. The main contributions of our work are the following:

\begin{itemize}
 \item The conception of a comprehensive, high-accuracy and high-precision measurement framework that enables us to measure the consumption of any type of device, as well as multiple heterogeneous devices synchronously.
 \item A rigorous methodology to perform whole-device energy measurements in laptops. As we will justify in detail later, we have chosen the laptop as target device in our measurements. Here we introduce the components of a common laptop, analyse which of them are key part of wireless transmissions, which others can become an issue (e.g., generating noise) and how to mitigate their impact.
 \item A collection of measurements and insights that deepens our understanding of the cross-factor.
\end{itemize}

The remainder of this paper is organized as follows. Section~\ref{sec:background} gives the current state-of-the-art on energy efficiency of wireless devices and establishes the foundation and motivation of this work. Section~\ref{sec:methods} presents our measurement framework: we describe in a reasoned manner the selected instrumentation, the testbed assembled for our experiments and its validation. Section~\ref{sec:components} assembles a measurement methodology by looking throughout the components of a common laptop. Section~\ref{sec:results} presents and discusses our results and its implications. Finally, Section~\ref{sec:conclusions} summarizes the conclusions of this paper.

\section{Background}\label{sec:background}

\subsection{Cross-factor: Towards a New Energy Model}

The seminal paper \cite{Feeney2001} gives the very first insight on energy consumption of wireless interfaces. This work was done on a per-frame basis by accounting for the energy drained by an 802.11 wireless card. Subsequent experimental works followed the same approach, which assumes that the network card dominates the consumption \cite{Jean-pierreEbert,Shih2002}. All these results show that the energy consumption of wireless transmissions/receptions can be characterised using a simple linear model. This fact arises from the three typical states of operation of a wireless card: idle, transmitting and receiving.

The model, commonly expressed in terms of power, has a fixed part $\rho_{id}$, device-dependent, attributable to the idle state, and it grows linearly with the airtime percentage $\tau$ (i.e., the fraction of time in which the card is transmitting or receiving).

This model has been assumed and widely used (implicitly or explicitly) in tens of papers \cite{Bruno2002,Carvalho2004,DajiQiao,Agrawal2004,Jyh-ChengChen2008,Garcia-Saavedra2011,Vaidya2002,Baiamonte2006,Zhang2012,Ergen2007,He2010,Baek2004,Sharma2009} to justify energy savings with optimizations of diverse kind: PHY layer rate and power \cite{DajiQiao}, MAC parameters \cite{Agrawal2004,Jyh-ChengChen2008,Garcia-Saavedra2011}, backoff operation \cite{Vaidya2002,Baiamonte2006}, idle state \cite{Zhang2012}, overhearing \cite{Ergen2007}, packet relying \cite{He2010}, data compression \cite{Baek2004,Sharma2009}, etcetera.

However, more recently, the novel work \cite{Serrano2014} performed extensive per-frame measurements for seven devices of different types (smartphones, tablets, embedded devices and wireless routers) and it unveiled that actually there is a non-negligible per-frame processing toll ascribed to the frame crossing the network stack. This component emerges as a new offset proportional to the frame generation rate, and it is not explained by the \emph{classical} model.

The \emph{new} model is expressed as follows:

\begin{equation}
 P(\tau, \lambda) = \underbracket{\rho_{id} + \rho_{tx}\tau_{tx} + \rho_{rx}\tau_{rx}}_{\text{classical model}} + \gamma_{xg}\lambda_{g} + \gamma_{xr}\lambda_{r}
\end{equation}

\noindent where:

\begin{description}[\IEEEsetlabelwidth{$\gamma_{xg}, \gamma_{xr}$]}\IEEEusemathlabelsep]
 \item [$\gamma_{xg}, \gamma_{xr}$] are the cross-factor for generation and reception respectively. It is the fixed energy toll, measured in mJ, that every frame pays while crossing the network stack.
 \item [$\lambda_{g}, \lambda_{r}$] are the frame generation and reception rates respectively.
\end{description}

The results obtained with this new multilinear model show that the so called cross-factor accounts for 50\% to 97\% of the per-frame energy consumption. Additional findings show that it is independent of the CPU load on some devices and \emph{almost} independent of the frame size.

Bearing in mind these results, it is imperative to reconsider existing schemes for energy efficiency in wireless communications. Old schemes like packet relying \cite{He2010} and data compression \cite{Baek2004,Sharma2009} may no longer be valid. On the contrary, packet batching or low-level packet generation could potentially produce real savings in this new component \cite{Serrano2014}.

\subsection{Motivation of this Work}

Due to its relevance, a better understanding of the cross-factor is required. We need to look at the kernel of the operating system, and specifically at the network stack, from an energy perspective: new tools and methodologies are needed to perform \emph{energy debugging} \cite{Pathak2011}.

Our intention is to separate the cross-factor into its constituent parts in order to comprehend the underlying causes, all this within the scope of providing an accurate mathematical description of the cross-factor which would give the energy model an unprecedented specificity. This would enable us to evaluate old energy efficiency strategies and to propose and test new schemes, both at device and at network level. Therefore, building a more general long-term measurement framework ---solid, accurate and flexible enough--- is required to cope the new challenge and those that will come.

Once this new framework is developed and validated, our proposal is to switch to a new target platform, more open and with easy access to in-kernel instrumentation. A laptop computer with a Linux-based distribution has all the resources to perform fine-grained energy debugging.

\section{A Comprehensive Measurement Framework for Wireless Devices}\label{sec:methods}

\begin{figure*}[t]
	\centering
		\subfloat[Testbed]{
			\label{fig:testbed} 
			\includegraphics[width=.38\linewidth]{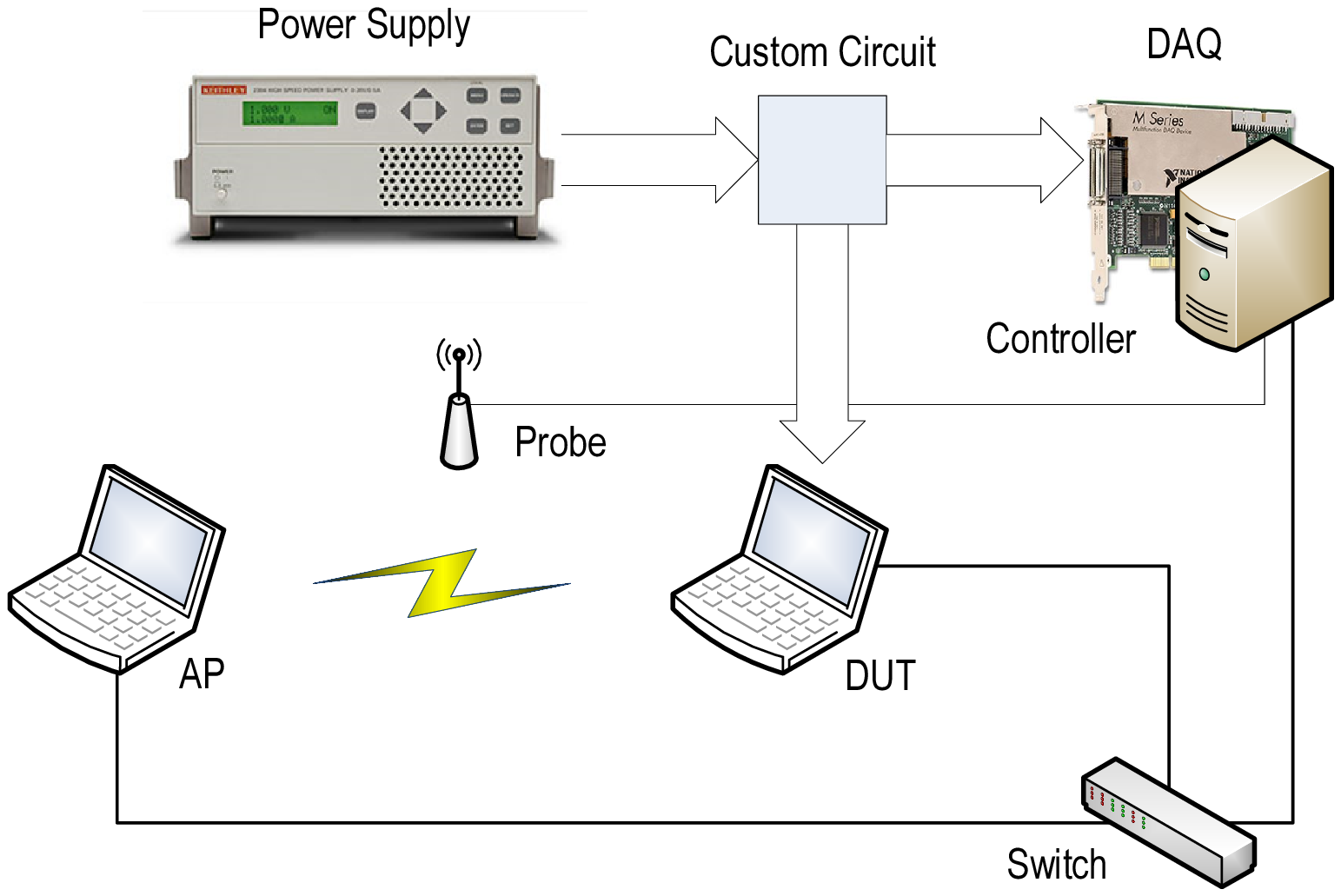}
		}
		\subfloat[Custom Circuit (simplified)]{
			\label{fig:circuito}
			\includegraphics[width=.30\linewidth]{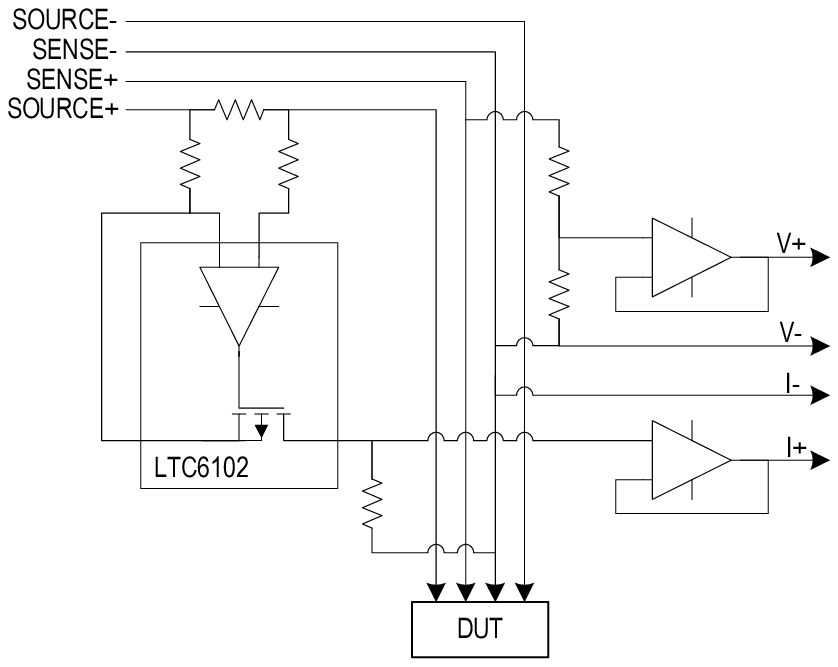}
		}
		\subfloat[Methodology]{
			\label{fig:secuencia}
			\includegraphics[width=.27\linewidth]{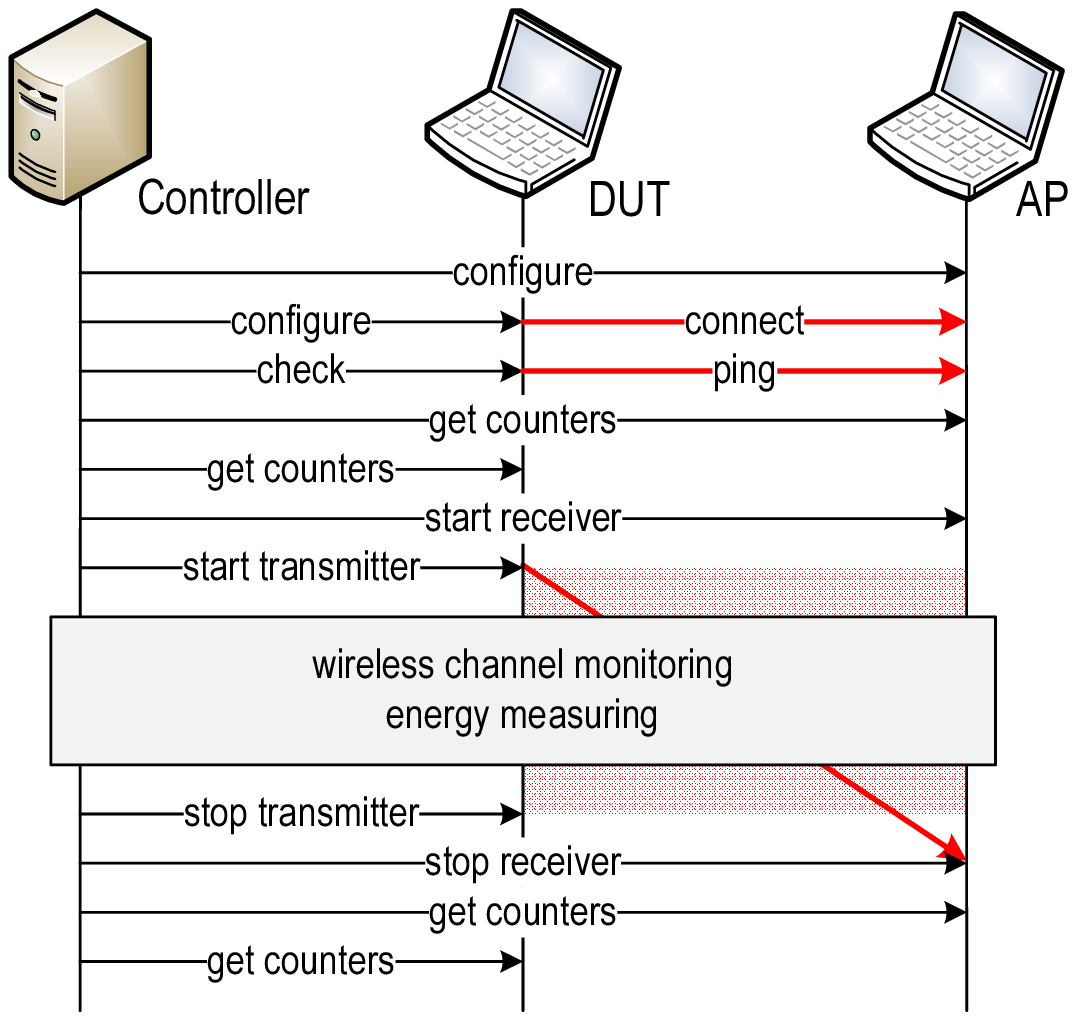}
		}
	\caption{(\ref{fig:testbed}) Testbed for energy measurements of a whole wireless device. (\ref{fig:circuito}) Simplified scheme of a custom circuit that connects the power supply with the DUT and extracts voltage and current signals accommodated to meet the DAQ input requirements. (\ref{fig:secuencia}) Time sequence of a single experiment.}
	\label{fig:methods}
\end{figure*}

The main specifications for our framework are the following:

\begin{itemize}
 \item High-accuracy, high-precision; in the range of mW.
 \item Avoid losing information between sampling periods, with events that last tens of microseconds.
 \item Wide range of devices, from low- to high-powered devices, while keeping accuracy and precision.
 \item Multiple devices under test (DUTs) at the same time, synchronously, for network-related energy measures (e.g., protocol/algorithm testing, energy optimization for a network as a whole). 
\end{itemize}

Based on the above, we selected the proper hardware components as we shall see. Making use of this framework, a testbed intended for deseeding the energy consumption of a Linux kernel stack is proposed and validated.

\subsection{Instrumentation}

\subsubsection{Hardware}

The most power-hungry laptops in the market rarely surpass the barrier of 100 W. For instance, typical Dell computers bring 65 or 90 W AC adapters with maximum voltages of 19.5 V. We selected the Keithley 2304A DC Power Supply, which is optimized for testing battery-operated, wireless communication devices such as smartphones, tablets or even laptops (up to 100 W, 20 V, 5 A) that undergo substantial load changes for very short time intervals. This power supply simulates a battery's response during a large load change by minimizing the maximum drop in voltage and recovering to within 100 mV of the original voltage in 40 us or less.

As measurement device, we selected the National Instruments PCI-6289, a high-accuracy multifunction data acquisition (DAQ) device.  It has 32 analogue inputs (16 differential or 32 single ended) with 7 input ranges optimized for 18-bit input accuracy, up to 625 kS/s single channel or 500 kS/s multi-channel (aggregate). The timing resolution is 50 ns with an accuracy of 50 ppm of sample rate.

A custom three-port circuit, specifically designed by our university's Technical Office in Electronics, converts the current signal to voltage and accommodates the voltage signal to the DAQ's input limits without precision loss. Fig.~\ref{fig:circuito} shows a simplified scheme of this circuit. The voltage drop in a small and high-precision resistor is amplified to measure the current signal. At the same time, a resistive divider couples the voltage signal. Considering that the DAQ card has certain settling time, it can be modelled as a small capacity which acts as a low pass filter. Thus, two buffers (voltage followers) are placed before the DAQ card to decrease the output impedance of the circuit \cite{ni2014}.

\subsubsection{Software}

A small command-line tool was developed to perform measurements on the DAQ card using the open-source Comedi\footnote{\url{http://comedi.org/}} drivers and libraries.

Regarding the in-kernel instrumentation, we take advantage of SystemTap\footnote{\url{https://sourceware.org/systemtap/}}, an open-source infrastructure around the Linux kernel that dramatically simplifies the gathering of information about a running Linux system (kernel, modules and applications). It provides a scripting language for writing instrumentation. SystemTap scripts are parsed into C code, compiled into a kernel module and hot-plugged into a live running kernel, eliminating the need for recompiling and rebooting.

\subsection{Testbed}

Fig.~\ref{fig:testbed} shows the proposed testbed. It is composed of two laptop computers ---the DUT and an access point (AP)--- and a controller. The controller is a workstation with the DAQ card installed and it performs the energy measurements. At the same time, it sends commands to the DUT and AP through a wired connection and monitors the wireless connection between DUT and AP through a probe.

The experimental methodology works as follows. Given a collection of parameter values (modulation coding scheme or MCS, transmission power, packet size, framerate), we run steady experiments for several seconds in order to gather averaged measures. Each experiment comprises the steps shown in Fig.~\ref{fig:secuencia}.

\begin{enumerate}
 \item AP and DUT are configured. The DUT connects to the wireless network created by the AP and checks the connectivity. Setting up this network in a clear channel is highly advisable to avoid interference. The  5 GHz band, with an 802.11a-capable card, has good candidates.
 \item The packet counters of the wireless interfaces are saved for later use.
 \item Receiver and transmitter are started. We use the \texttt{mgen}\footnote{\url{http://cs.itd.nrl.navy.mil/work/mgen/}} traffic generator and a simple \texttt{netcat} at the receiver.
 \item The controller monitors the wireless channel and collects an energy trace that will be averaged later.
 \item Transmitter and Receiver are stopped.
 \item Because of the unreliability of the wireless medium, the packet counters, together with the monitoring information, are used to ensure that the experiment was successful (i.e., the traffic seen agrees with the configured parameters).
\end{enumerate}

\subsection{Validation}

In order to validate our measurement framework, several experiments were performed with one of the devices studied in \cite{Serrano2014} as DUT. We selected the Soekris net4826-48 equipped with an Atheros AR5414-based 802.11a/b/g Mini-PCI card because it is the one with the largest cross-factor. The OS was Linux Voyage with kernel 2.6.30 and the MadWifi driver v0.9.4.

The first task was to perform the energy breakdown given in \cite{Serrano2014} in transmission mode:

\begin{itemize}
 \item \textbf{User space}: The Soekris generates packets using \texttt{mgen}, but they are discarded before being delivered to the OS, by using the \texttt{sink} device rather than \texttt{udp}.
 \item \textbf{Kernel space}: Packets cross the network stack and are discarded in the driver, by commenting the \texttt{hardstart} MadWifi command that performs the actual delivery of the frame to the wireless network interface card (NIC).
 \item \textbf{Wireless NIC}: Packets are transmitted, i.e., are delivered to the wireless medium.
\end{itemize}

The NoACK functionality of 802.11e was activated in order to avoid ACK receptions. Thus, the energy model that governs Soekris' complete transmissions is simplified as follows:

\begin{equation}\label{ec:validation1}
 P(\tau, \lambda) = \rho_{id} + \rho_{tx}\tau + \gamma_{xg}\lambda
\end{equation}

\begin{figure}[t]
	\centering
	\includegraphics[width=\linewidth]{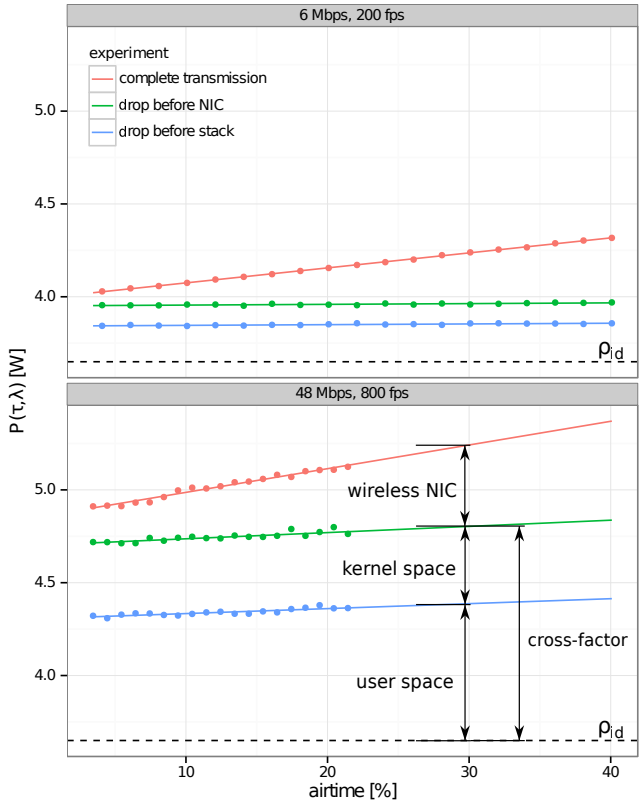}
	\caption{Power consumption breakdown versus airtime in a Soekris net4826-48.}
	\label{fig:validation1}
\end{figure}

Fig.~\ref{fig:validation1} represents the Equation~(\ref{ec:validation1}) (red lines) and depicts how the energy toll splits across the processing chain with different parameters (blue and green lines). The dashed line depicts the idle consumption as a reference, $\rho_{id}=3.65(1)$ W.

Indeed, these results are quite similar to \cite{Serrano2014} and confirm that the cross-factor accounts for the largest part of the energy consumption. Moreover, \cite{Serrano2014} reports that the cross-factor is \emph{almost} independent of the packet size. Interestingly, our results have captured a small dependence that can be especially observed in the 600 fps case.

\begin{figure}[t]
	\centering
	\includegraphics[width=\linewidth]{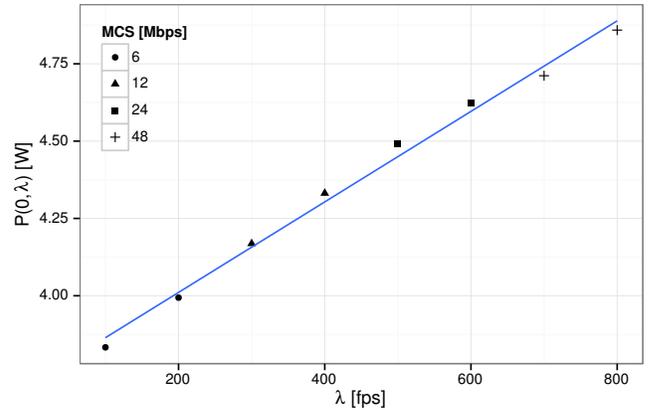}
	\caption{Power consumption offset ($\tau=0$) versus framerate in a Soekris net4826-48.}
	\label{fig:validation2}
\end{figure}

Finally, we can derive the cross-factor value and compare it. Taking the offset of the red regression lines of Fig.~\ref{fig:validation1}, we can plot Fig.~\ref{fig:validation2} and fit the points to the Equation~(\ref{ec:validation1}) with $\tau=0$. This regression yields the values $\rho_{id}=3.72(4)$~W ($3.65(1)$~W measured) and $\gamma_{xg}=1.46(7)$~mJ, quite close to the values reported in \cite{Serrano2014}.

\section{Components of a Common Laptop}\label{sec:components}

As pointed out previously in Sections~\ref{sec:intro}-\ref{sec:background}, we have chosen the laptop as DUT in our measurements using the testbed described and validated in Section~\ref{sec:methods}. The main reason is that this platform is flexible and powerful enough to provide us with the proper debugging tools.

However, a laptop is not without drawbacks as a complex and power-hungry piece of hardware. It comprises a number of components, both hardware (battery, screen, hard disk drive, fan, wireless card, RAM memory, CPU) and software (services, kernel, drivers), that require a thorough discussion for one of two reasons: 1) they are not present in other devices or 2) they are essentially different.

The laptop selected for our experiments is a Dell Latitude E5540 with Intel Core i5-4300U CPU at 1.9 GHz and 8 GB of SODIMM DDR3 RAM at 1.6 GHz, equipped with an Atheros AR9280-based 802.11a/b/g/n Half Mini-PCI Express card and running Fedora Linux 20.

\subsection{Battery}

The battery is a serious obstacle for energy measurements. Although using it as power source is actually possible, it is totally impractical because it prevents long term experiments, and the constant need for recharging is a waste of time. Then, the use of an external power source is highly advisable, but in this case the battery must be removed to avoid noise coming from battery charging and discharging.

Nevertheless, supplying DC voltage through the power jack socket is not enough. Most laptops are capable of detecting the AC adapter. This is done through a third connection in the power jack. In the case of Dell computers, this third wire goes to a transistor-shaped component placed in the AC transformer. Actually, this component is a small memory that can be read using a 1-wire protocol. It stores a serial number identifying the AC adapter.

If the laptop does not detect this memory, the BIOS can do improper things. For instance, we have detected that the BIOS of Dell computers do not allow the OS to control CPU frequency scaling. Fortunately, it is very straightforward to borrow such component from an official AC adapter and attach it permanently to the third connection of the power jack socket.

\subsection{Screen}

Same as for smartphones \cite{Carroll2010}, the laptop screen is the most energy-hungry device. It accounts for more than a half of the energy consumed by our laptop when it is just powered on and idling. Thus, the screen constitutes a very high and variable (as it depends on the GPU activity) baseline consumption that must be avoided in wireless experiments.

In Linux, this can be done by finding the backlight device entry in the \texttt{/sys/class} subsystem and simply resetting it. The screen goes off.

\subsection{Hard Disk Drive}

Regarding the system's non-volatile memory, we cannot get rid of it because it is needed for the OS storage. Commonly, laptops carry hard disk drives (HDD), which are mechanical devices powered with voltages ranging from 5 to 12 V. HDDs are proven to be energy-hungry devices with a consumption variability in ascending order of tens of W \cite{Hylick2008}. As a consequence, every read/write during an experiment generates an intractable noise.

On the contrary, all the devices studied in \cite{Serrano2014} use flash memories. This kind of non-volatile memory is the best option, because its consumption variability is three orders of magnitude below HDD's \cite{Grupp2009}. In our experiments, we replaced the original HDD by a solid-state disk (SSD). As an SSD is composed of NAND flash units, its consumption is far more stable.

\subsection{Fan}

The thermal characteristics of a laptop computer require a cooling subsystem: heat sinks (CPU and GPU), air ducts and one fan (at least). The fan is regulated dynamically with a pulse-width modulation (PWM) technique. This component becomes an unpredictable source of electrical noise, as its operation point depends on the computer's thermal state. 

Suppressing the fan is not an option because, at some point, the CPU will heat and the computer will turn off. Our solution was to set it at fixed medium speed with the help of the \texttt{i8k} kernel module and the \texttt{i8kfan} user-space application.

\subsection{Wireless Card}

This is the last part of the wireless transmission chain and the first of the reception one. In principle, the energy models reviewed in Section~\ref{sec:background} assure us that a linear behaviour is expected, independently of the manufacturer or model. However, there are a couple of factor may lead us to select a given card.

\begin{itemize}
 \item \textbf{Capabilities}: Nowadays, it is very difficult to perform interference-free wireless experiments over ISM bands without an anechoic chamber (especially when all your fellows are within the same research topic). The 2.4 GHz band is typically overcrowded, while in the 5 GHz band we have better chances to find a clear channel. Thus, an 802.11a-capable card is advisable. 
 \item \textbf{Manufacturer}: Distinct manufacturers (and models) have better or worse driver support in the Linux kernel. For instance, Intel PRO/Wireless cards are known for requiring a binary firmware to operate. On the other hand, Atheros released some source from their binary HAL to help the open-source community add support for their chips. As a result, there are completely free and open-source drivers available for all Atheros chipsets.
\end{itemize}

The factory default card of our Dell Latitude was an Intel PRO/wireless card, and it was replaced by an Atheros AR9280-based 802.11a/b/g/n Half Mini-PCI Express.

\subsection{RAM Memory}

The random-access memory is a fundamental peripheral device in a computer system: it holds the instructions of the running programs ---the kernel of the OS included--- and the data associated. Therefore, our first guess was that the RAM memory could play a meaningful role in the energy consumption of a wireless communication.

\subsection{CPU}

\begin{figure}[t]
	\centering
	\includegraphics[width=.49\linewidth]{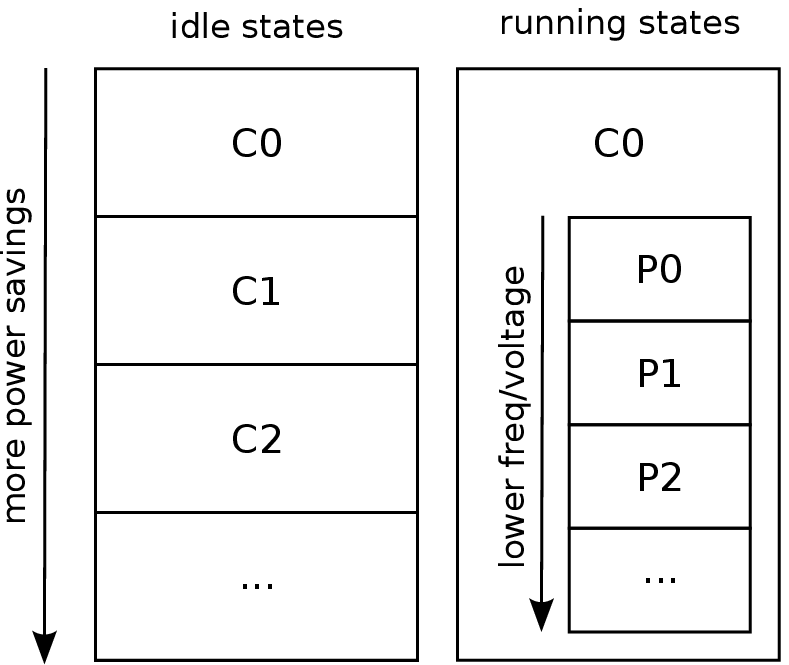}
	\caption{CPU P- and C-states.}
	\label{fig:states}
\end{figure}

The CPU is another power-hungry component. For many years, the first CPUs were like bulbs: they were consuming the same power whether doing something useful or not. In fact, they executed \emph{junk code} (i.e., a loop of NOPs) in idle time. Later, CPU architects realised that more intelligent things can be done in such periods of time. For instance, performing some kind of energy saving mechanism.

Nowadays, CPUs are becoming more and more complex. Without seeking to be exhaustive, a modern CPU has arithmetic control units (ALUs), pipelines, a control unit, registers, several levels of cache, some clocks, etcetera. But more interestingly, modern CPUs implement several power management mechanisms (see Fig.~\ref{fig:states}) covered by the Advanced Configuration and Power Interface (ACPI).

\begin{itemize}
 \item \textbf{P-states}: Also known as \emph{frequency scaling}. When the CPU is running (i.e., executing instructions), one of these states apply. P0 means maximum power and frequency. As Px increases, the voltage is lower and, thus, the frequency of the clock, and thus the energy consumed, is scaled down.
 \item \textbf{C-states}: When the CPU is idle, it enters a Cx state. The C0 means maximum power, because junk code is being executed. This can be a little bit confusing because, as Fig.~\ref{fig:states} shows, the CPU is also in C0 when running. In general, C0 means \emph{the CPU is busy doing something}, whether executing actual programs (running) or something not useful (idle). In C1, the CPU is halted, and can return to an executing state near instantaneously. In C2, the main clock is stopped and, as Cx increases, more and more functional units are shut off. As a consequence, returning from a deep C-state is very expensive in terms of latency.
\end{itemize}

Finally, multicore systems introduce additional complexity. As a clue, the OS decides how many cores become active at any time, and each core has its own power management subsystem (i.e., P- and C-states). Therefore, we work always in single core mode to simplify the analysis.

\subsection{Services}

There may be a lot of active user-space services (also called \emph{daemons}) in a Linux system by default. They can add noise to our measurements in two ways: by consuming CPU time and writing logs to disk. Hence, disabling not essential services is desirable.

\subsection{Kernel}

There exist two power management subsystems for each CPU in the Linux kernel: \texttt{cpufreq}\footnote{\url{https://www.kernel.org/doc/Documentation/cpu-freq/}} controls P-states and \texttt{cpuidle}\footnote{\url{https://www.kernel.org/doc/Documentation/cpuidle/}} controls C-states. Both subsystems have the same architecture, separating mechanism (driver) from policy (governor).

\begin{itemize}
 \item \textbf{Driver}: It provides the platform-dependent state detection capability and the mechanisms to support entry/exit into/from different states. By default, there exists an ACPI driver that implements standard APIs. Usually, CPU-specific drivers are capable of detecting more states than ACPI-compliant ones.
 \item \textbf{Governor}: It is an algorithm that takes in several system parameters as input and decides the state to activate.
\end{itemize}

The \texttt{cpufreq} governor has several policies that focus on certain P-states or frequencies to the detriment of others, e.g., \texttt{performance} (high frequencies) or \texttt{powersave} (low frequencies). It is also possible to manually fix certain frequency or range of frequencies.

The \texttt{cpuidle} governor takes in the next timer event as main input. Each C-state has a certain energy cost and exit latency. Thus, intuitively there are two decision factors that the \emph{menu} governor (the most common) must consider: the energy break-even point and the performance impact. The next timer event is a good predictor in many cases, but not perfect since there are other sources of wake-ups (e.g., interrupts). Therefore, it computes a correction factor using an array of 12 independent factors through a running average. Moreover, it is possible to manually disable the C-states (excepting C0).

\subsection{Drivers}

All Linux drivers are compiled as separate modules. In particular, wireless drivers\footnote{\url{http://wireless.kernel.org/}}, along with the entire 802.11 subsystem, can be compiled out-of-tree within the \emph{backports} project\footnote{\url{https://backports.wiki.kernel.org}}. This is very useful in order to use latest drivers on older kernels.

The wireless driver module interacts directly with the NIC. In our case, the selected card uses the \texttt{ath9k} driver. The function \texttt{ath9k\_tx()} is the entry point for the transmission path. The driver fills the transmission descriptors, copies the buffer into the NIC memory and sets up several registers that trigger the transmission.

In \cite{Serrano2014}, the authors claim that their methodology discarded packets \emph{right after the driver} in order to perform the energy breakdown depicted in Fig.~\ref{fig:validation1}. This statement becomes uncertain when a closer look at any wireless driver is taken. Discarding a frame \emph{before} the buffer is copied into the NIC implies that \emph{only half} of the driver is actually taken into account for the cross-factor value. On the other hand, if we try to discard it in the very last instruction of the driver (i.e., avoiding setting the register that triggers the hardware transmission), then the module crashes. This occurs because the buffer is already in the NIC's memory and it needs to be cleaned up, a non-trivial task that anyway would consume more energy.

This, combined to the fact that drivers differ greatly from one to another, makes it not advisable to include the driver into the definition of cross-factor, given the difficulty of isolating driver and NIC consumptions.

As with the variety of drivers mentioned above, a similar argument can be wield against the user-space consumption. Therefore and from here on, we define \emph{kernel cross-factor} as the energy consumed from the system call that delivers the message until the driver is reached. \emph{Cross-factor}, as is, maintains the definition given in \cite{Serrano2014} to avoid confusion.

We would also like to highlight that our way of interrupting the transmission path by discarding a frame in the driver is a bit different from \cite{Serrano2014}. They conduct this breakdown commenting a driver function. This method implies the need for recompiling the driver, which is time-consuming and not very portable (think about a similar task inside the kernel core).

For our part, we look for a short string in the beginning of packet payloads. The presence of this \emph{magic string} triggers the packet drop. Our method, despite introducing a very little overhead, is agile and portable (for instance, it can be implemented on the fly using SystemTap).

\section{Networking Measures on a Laptop}\label{sec:results}

In our measurements, we use Fedora-default pre-compiled kernels. We arranged two separate partitions: one with Fedora 12 and kernel version 2.6.32 and the other with Fedora 20 and kernel 3.14. Only the latter supports Intel-specific drivers, thus we use ACPI drivers only in order to operate under similar conditions.

Intel Haswell processors support up to eight C-states: C1, C1E, C3, C6, C7s, C8, C9, C10. However, the BIOS reports two C-states to the ACPI driver, which are named C1 and C2. We have verified by comparing idle consumptions of each C-state that the correspondence is as follows:

\begin{itemize}
 \item ACPI C1 $=$ Intel C1: the CPU is halted and stops executing instructions when it enters into idle mode.
 \item ACPI C2 $=$ Intel C6: this is a new sleep state introduced in the Haswell architecture.
\end{itemize}

\subsection{Cross-factor: Separating the Wheat from the Chaff}

\begin{figure}[t]
	\centering
	\includegraphics[width=\linewidth]{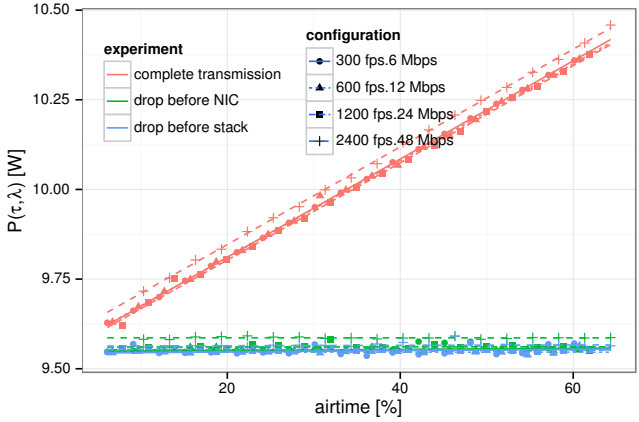}
	\caption{Energy breakdown with the CPU fixed at P0-C0 states.}
	\label{fig:c0}
\end{figure}

We have identified the two components suspected of being responsible of the cross-factor, and our first priority is to quantify their impact:

\begin{itemize}
 \item \textbf{CPU}: As the cross-factor is caused by software processing, the CPU is expected to be the main source of energy drain.
 \item \textbf{RAM Memory}: It stores the instructions to be executed as well as the associated data.
\end{itemize}

It is not possible to regulate the activity of the RAM memory, but it can be done with the CPU. For this purpose, the CPU was fixed at P0-C0 states, i.e., always running at maximum frequency, maximum energy consumption. We performed an energy breakdown using this configuration in kernel 3.14 and the results are shown in Fig.~\ref{fig:c0}.

The lines appear superimposed: the laptop is consuming the same power among different parameters, different packet rates. Hence, one important conclusion to be drawn is that the RAM memory has no significant impact in the overall energy consumption of wireless transmissions. The noise can be ascribed to the fact that not all the instructions consume exactly the same energy \cite{Tiwari1996}. Other possible sources of noise are cache and pipeline flushes.

With this simple experiment, we have demonstrated that the CPU is the leading cause of cross-factor in laptops, and it is clear that the \texttt{cpuidle} subsystem has a central role, because a CPU spends most of the time in idle mode \cite{Barroso2007}. From now on, and in order to take a deeper look at C-states, we remove a variable by keeping the P-state fixed at P0 (maximum frequency).

\subsection{Power Consumption in Unattended Idle Mode}

\begin{figure}[t]
	\centering
	\includegraphics[width=\linewidth]{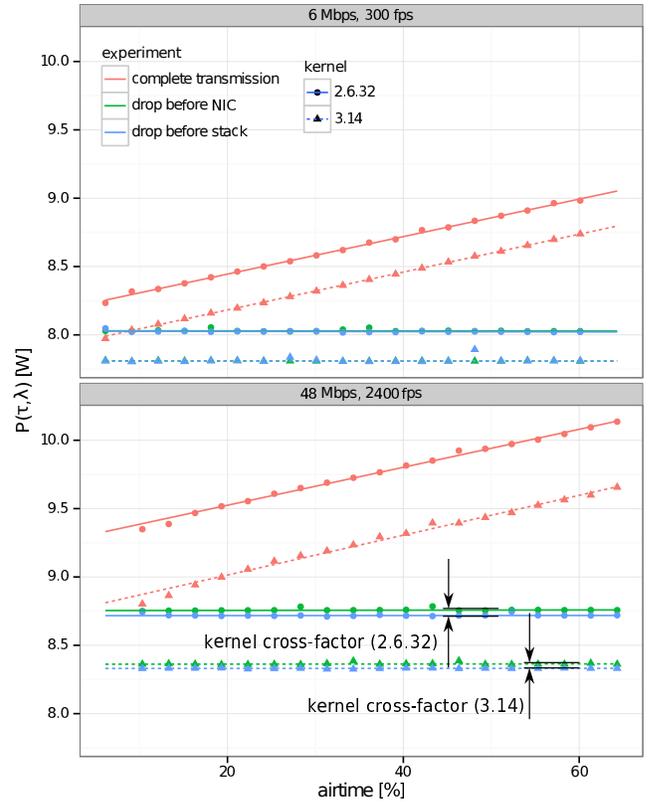}
	\caption{Power consumption breakdown versus airtime with fixed C1 state for kernels 2.6.32 and 3.14.}
	\label{fig:c1}
\end{figure}

\begin{figure}[t]
	\centering
	\includegraphics[width=\linewidth]{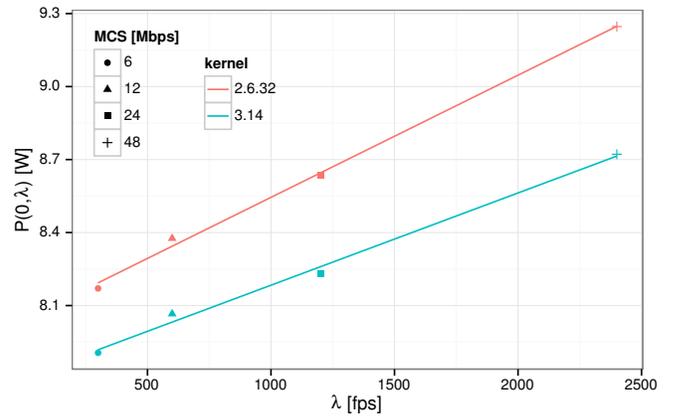}
	\caption{Power consumption offset ($\tau=0$) versus framerate with fixed C1 state for kernels 2.6.32 and 3.14.}
	\label{fig:c1-cfactor}
\end{figure}

The Soekris net4826-48 is equipped with an AMD Geode SC1100 CPU that supports ACPI C1, C2 and C3 states\footnote{\url{http://datasheets.chipdb.org/upload/National/SC1100.pdf}}. Unfortunately, it seems that Linux distributions for embed devices, such as Voyage Linux, disable \texttt{cpuidle} in their kernels, which means that the OS has no control over the idle mode. In such conditions, we know now that the CPU cannot be in C0 all the time, because the device does not consume the same power with different parameters. What is happening then?

Back to our laptop, it is possible to disable \texttt{cpuidle} through the kernel command-line. The idle power consumption in this situation, which we call \emph{unattended idle mode}, reveals that the laptop is entering C1. This fact can be extrapolated to the Soekris case, which makes sense, since there is no governor to resolve what C-state is the more suitable. Thus, the processor simply halts when there is no work to do.

Fig.~\ref{fig:c1} shows the energy breakdown for both kernels, 2.6.32 and 3.14, when the C1 state only is enabled. The obtained kernel cross-factor is almost negligible, which suggests that Intel Haswell's C1 state saves a very small amount of power, unlike the Soekris' C1 state as shown in Fig.~\ref{fig:validation1}.

There is also a baseline power difference between kernels. This offset can be ascribed to several factors. For instance, a lot of code has changed ---and probably improved--- between those kernel versions. In particular, the scheduler and the \texttt{cpuidle} algorithms have evolved. Moreover, the compiler used has changed also.

At this respect, we can calculate the complete cross-factor (including the user-space, as done in Section~\ref{sec:methods}) by extracting the slopes of the regressions of Fig.~\ref{fig:c1-cfactor}. These values are comparable to the Linksys case reported in \cite{Serrano2014}: $0.51(2)$ mJ (kernel 2.6.32) and $0.38(2)$ mJ (kernel 3.14).

It is also important to note that, unlike results from Fig.~\ref{fig:validation1}, there is absolutely no dependence on the frame size in this case. Our guess is that RAM memory consumption would be proportional to the frame size and may have a small but still perceptible impact in low-power devices, but it is negligible compared to a laptop's CPU consumption. As a consequence, the frame size can be removed as a parameter from the cross-factor analysis in laptops.

\subsection{Power Consumption with Full \texttt{cpuidle} Subsystem}

\begin{figure}[t]
	\centering
	\includegraphics[width=\linewidth]{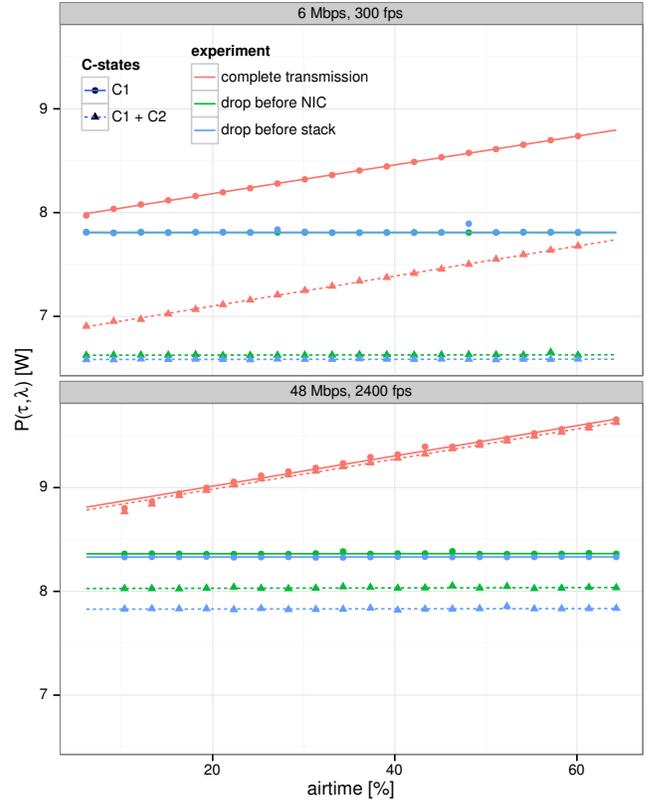}
	\caption{Power consumption breakdown versus airtime with two \texttt{cpuidle} configurations for kernel 3.14.}
	\label{fig:c12}
\end{figure}

\begin{figure}[t]
	\centering
	\includegraphics[width=\linewidth]{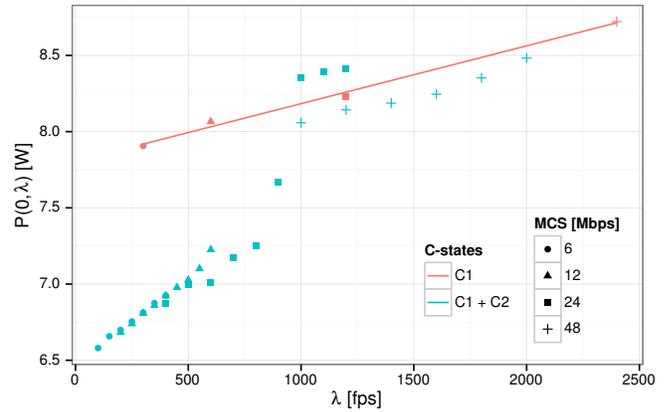}
	\caption{Power consumption offset ($\tau=0$) versus framerate with two \texttt{cpuidle} configurations for kernel 3.14.}
	\label{fig:c12-cfactor}
\end{figure}

With the knowledge acquired so far, we can move onto a more realistic scenario by enabling the whole \texttt{cpuidle} subsystem, i.e., keeping both ACPI C-states enabled and letting the governor decide.

Fig.~\ref{fig:c12} depicts the energy breakdown for kernel 3.14 with full \texttt{cpuidle} subsystem (C1+C2 enabled) and compares it to the previous case (C1 only). By enabling C2, the consumption appears to be always lower up to driver level (blue and green lines). Nevertheless, the consumption of complete transmissions (red lines) is lower in the 300 fps case, but it is the same in the 2400 fps case.

Fig.~\ref{fig:c12-cfactor} compares the offsets of complete transmissions in Fig.~\ref{fig:c12} for both cases: C1+C2 and C1 only. The red line corresponds to C1: as expected, its behaviour is linear as seen in Fig.~\ref{fig:c1-cfactor}. On the other hand, the C1+C2 case (blue points) is not linear globally. It comprises three clearly distinct parts: when the framerate is low, there is an approximately linear behaviour because the CPU only uses C2; when the framerate is high, C2 is no longer used, and the slope matches the red line; between them, the behaviour becomes unpredictable because of the mix of C1 and C2. Therefore, the cross-factor as defined in \cite{Serrano2014} makes no sense anymore. When all C-states are active, there is no more linear behaviour: we cannot talk neither about a slope nor a fixed energy toll per frame.

Furthermore, we had assumed, as \cite{Serrano2014}, that we can simply drop the packets at certain points, measure the mean power up to those points and represent all this as an energy breakdown. But obviously this is not true either. For instance, the last plot of Fig.~\ref{fig:c12} (48 Mbps, 2400 fps) shows that the CPU is not entering C2 when complete transmissions are performed, as the consumption is the same as the C1-only case. On the other hand, the CPU is clearly spending some time in C2 when the frames are dropped early. Even it seems that the network stack is consuming more power because the energy band (between green and blue lines) is larger. Evidently, it should be the opposite: the stack would be consuming less power as soon as it enters a lower C-state.

\subsection{Exploring the \texttt{cpuidle} Subsystem}

\begin{figure}[t]
	\centering
	\includegraphics[width=\linewidth]{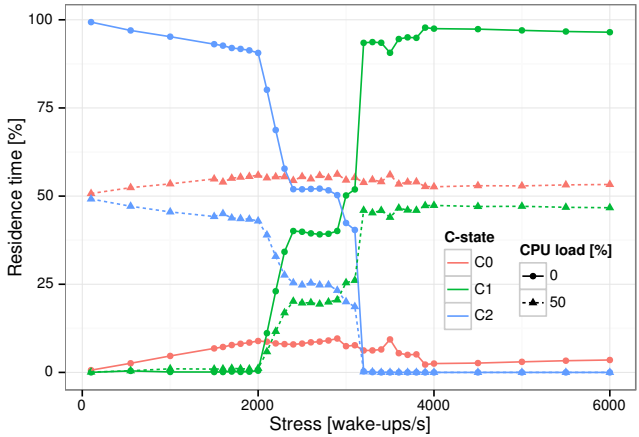}
	\caption{Residence time of each C-state versus wake-ups/s for kernel 3.14. Each wake-up does nothing.}
	\label{fig:residency-wakeups}
\end{figure}
\begin{figure}[t]
	\centering
	\includegraphics[width=\linewidth]{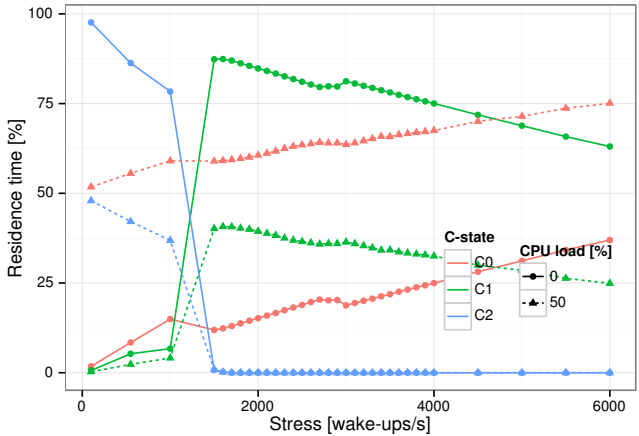}
	\caption{Residence time of each C-state versus wake-ups/s for kernel 3.14. Each wake-up performs a UDP transmission.}
	\label{fig:residency-wakeups-udp}
\end{figure}
\begin{figure}[t]
	\centering
	\includegraphics[width=.95\linewidth]{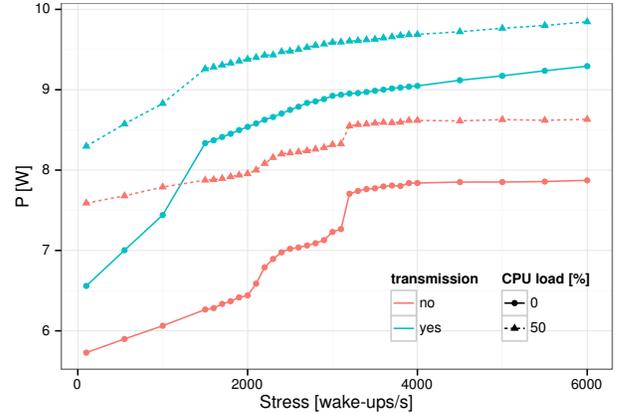}
	\caption{Power consumption offset versus wake-ups/s for kernel 3.14. A comparison between the "wake-up" (Fig.~\ref{fig:residency-wakeups}) and "wake-up + UDP transmission" (Fig.~\ref{fig:residency-wakeups-udp}) cases is made.}
	\label{fig:power-wakeups}
\end{figure}

As stated in Section~\ref{sec:components}, the \texttt{cpuidle} subsystem is a very complex component. Kernel timer events are the main input for the governor algorithm as they often indicate the next wake-up of the CPU, but the running average used to scale the latter makes it unpredictable, as it depends on the recent state of the whole machine. The purpose of this section is to shed light on the linkage between the residence time of C-states, the number of wake-ups per second, the CPU load and the transmission of wireless frames.

We implemented a very simple application with two modes of operation: it is capable of setting a kernel timer at a given constant rate and, when this timer is triggered, it 1) does nothing or 2) sends a UDP packet. At the same time, it calculates the mean residence time of each C-state over the whole execution. Figs.~\ref{fig:residency-wakeups}-\ref{fig:residency-wakeups-udp} have been compiled using this tool. The additional CPU load was added on top of the latter using a modified version of \texttt{lookbusy}\footnote{\url{http://www.devin.com/lookbusy/}}. Fig.~\ref{fig:power-wakeups} compares the two previous figures in terms of power consumption.

In Fig.~\ref{fig:residency-wakeups}, the only source of wake-ups is the kernel timer that our tool sets. Each C-state is represented by a colour, and shapes and line types distinguish between CPU loads. The first observation is that the addition of a substantial source of CPU load has no impact on the distribution of residence times. Another important clue is that, up to 2000 and from 3500 wake-ups/s onwards, there is only one active idle state ---C2 or C1 respectively (C0 means executing)---, and the behaviour is linear. This fact can be verified by checking the power consumption (Fig.~\ref{fig:power-wakeups}, red lines). From 2000 to 3500 wake-ups/s, the transition between C-states occurs in a non-linear way.

In Fig.~\ref{fig:residency-wakeups-udp}, on other hand, there is another source of wake-ups: hardware interrupts caused by the wireless card each time a packet is sent. The transition between states occurs earlier because there is actually twice the wake-ups. And, again, CPU load shows no impact on the distribution of residence times.

These are partial results and are limited to constant rate wake-ups, but these findings are in line with the non-linearities previously discovered in the cross-factor and they confirms the enormous complexity we face.

\section{Conclusions}\label{sec:conclusions}

This paper follows the path set out by \cite{Serrano2014} with the discovery of the cross-factor, an energy toll not accounted by classical energy models and associated to the very fact that frames are processed along the network stack. On the achievement of this goal, we have built and validated a comprehensive, high-accuracy and high-precision measurement framework capable of measuring any type of device, as well as multiple heterogeneous devices synchronously. We have introduced the laptop as a more suitable device to perform whole-device energy measurements in order to deseed the root causes of the cross-factor by taking advantage of the wide range of debugging tools that such platform enables.

Our results, albeit preliminary, provide several fundamental insights on this matter:

\begin{itemize}
 \item We have identified the CPU as the leading cause of the cross-factor in laptops. Thus, the cross-factor shows absolutely no dependence on the frame size, because the RAM memory has no significant impact in the overall energy consumption of wireless transmissions. On the other hand, low-powered devices, like the Soekris, show a very small but perceptible dependency that can be ascribed to the RAM memory.
 \item The CPU's C-state management plays a central role in the energy consumption, because a CPU spends most of the time in idle mode.
 \item When the C-state management subsystem is not present in the OS, the device enters C1 in idle mode (halted) and cannot benefit from lower idle states.
 \item In contrast to low-powered devices, the C1 state of a laptop's CPU saves a very small amount of power.
 \item With a fully functional C-state management subsystem, the linear behaviour disappears. In consequence, we cannot talk about cross-factor as a fixed energy toll per frame.
 \item A non-linear behaviour implies that we cannot perform energy breakdowns by dropping packets inside the transmission chain. Therefore, new methodologies and techniques are required to enable energy debugging.
 \item C-state residence times depend primarily on the number of wake-ups per second produced by software and hardware interrupts. However, they show no dependence on the CPU load.
\end{itemize}

Further research is needed in order to fully understand the key role of the C-state subsystem in the energy consumption of wireless communications, as well as to investigate other processor capabilities not accounted for here, such as P-states and multicore support.

\balance
\bibliographystyle{IEEEtran}
\bibliography{paper}
\end{document}